\documentclass[12pt,psfig,epsf]{article}
\usepackage{psfig}

\renewcommand{\theequation}
{\arabic{section}.\arabic{equation}}

\makeatletter
\def\eqnarray{ \stepcounter{equation} \let\@currentlabel=\theequation
 \global\@eqnswtrue
 \global\@eqcnt\z@
 \tabskip\@centering
 \let\\=\@eqncr
 $$\halign to \displaywidth\bgroup\@eqnsel\hskip\@centering
 $\displaystyle\tabskip\z@{##}$&\global\@eqcnt\@ne
 \hfil$\displaystyle{{}##{}}$\hfil
 &\global\@eqcnt\tw@$\displaystyle\tabskip\z@{##}$\hfil
 \tabskip\@centering&\llap{##}\tabskip\z@\cr}
\makeatother

\makeatletter
\def\@arrayacol{\edef\@preamble{\@preamble \hskip .5\arraycolsep}}
\def\array{\let\@acol\@arrayacol \let\@classz\@arrayclassz
\let\@classiv\@arrayclassiv \let\\\@arraycr\def\@halignto{}\@tabarray}
\makeatother

\makeatletter
\newcounter{subeqncnt}
\def\thesubeqncnt{\alph{subeqncnt}}
\def\subequations{\begingroup%
   \stepcounter{equation}\edef\@tempa{\theequation}%
   \let\c@equation\c@subeqncnt\c@subeqncnt\z@
   \edef\theequation{\@tempa\noexpand\thesubeqncnt}}

\makeatother

\newcommand{\be}{\begin{equation}}
\newcommand{\ee}{\end{equation}}

\newcommand{\beqa}{\begin{eqnarray}}
\newcommand{\eeqa}{\end{eqnarray}}
\newcommand{\nn}{\nonumber}

\newcommand{\eqref}[1]{(\ref{#1})}

\def\CL {{\cal L}}
\def\CM {{\cal M}}

\begin{document}

\setlength{\baselineskip}{7mm}
\begin{titlepage}
\begin{flushright}
{\tt NRCPS-HE-02-08} \\

February, 2008
\end{flushright}

\vspace{1cm}

\begin{center}
{\it \Large Tensor Gauge Boson Production    \\ in  \\
 High  Energy Collisions
}

\vspace{1cm}

{ \it{Spyros Konitopoulos,} }
\it{Raffaele Fazio}\footnote{On a leave of absence from
Departamento de Fisica, Universidad Nacional de Colombia,
Bogot´a.}
and
{ \it{George  Savvidy  } }

\vspace{0.5cm}

 {\it Institute of Nuclear Physics,} \\
{\it Demokritos National Research Center }\\
{\it Agia Paraskevi, GR-15310 Athens, Greece}

\end{center}

\vspace{1cm}

\begin{abstract}

\end{abstract}
We calculated the leading-order cross section for the helicity two
tensor gauge bosons production in fermion pair annihilation process.
We compare this cross section with a similar
annihilation processes in QED with two photons in
the final state and with two gluons in QCD.

\end{titlepage}

\pagestyle{plain}

\section{\it Introduction}

Our intention in this article is to calculate leading-order differential
cross section for the tensor gauge bosons production in the fermion pair
annihilation process. The process is illustrated in   Fig.\ref{fig1}. and
receives contribution from three Feynman diagrams shown in  Fig.\ref{fig3}.
This diagrams are similar to the QED and QCD diagrams for the annihilation processes
with two photons or two gluons in the final state. The difference between these
processes is in the
actual expressions for the corresponding interaction vertices. The corresponding
vertices for the tensor bosons can be found through the extension of the gauge principle \cite{Savvidy:2005fi}.
The extended gauge principle allows to define a gauge invariant Lagrangian $\CL$ for
high-rank tensor gauge fields and their cubic and quartic interaction vertices
\cite{Savvidy:2005fi,Savvidy:2005zm,Savvidy:2005ki}:
$$
{{\cal L}} =  \CL_{YM}~+~   \CL_{2}~+ ~\CL^{'} _{2}...
$$
The Feynman rules for this Lagrangian can be derived from the functional
integral over the fermion fields $\psi_{i },~\psi^{\mu}_{i },~
...$ and over the gauge boson fields
$A^{a}_{\mu},~A^{a}_{\mu\nu},...$.

Not much is known about physical properties of such gauge
field theories
\cite{fierz,fierzpauli,schwinger,fronsdal,Bengtsson:1983pd,Witten:1985cc,Metsaev:2007rn}
and in the present article we shall ignore subtle aspects (ghosts)
of functional integral quantization procedure because we limited
ourselves to calculating only leading-order tree diagrams.
Expanding the functional integral in perturbation theory, starting with the free
Lagrangian, at $g=0$, one can see that free theory contains tensor gauge bosons
and fermions of different spins  with
cubic and quartic interaction vertices \cite{Savvidy:2005fi,Savvidy:2005zm,Savvidy:2005ki}.
Explicit form of these vertices is presented in \cite{Savvidy:2005ki}.

In the next section we shall present the Feynman diagrams for the given process,
the expressions for the corresponding vertices, the transition matrix element and
unpolarized cross section in the center-of-mass frame.
In the third section we shall check that the transition amplitude is gauge invariant,
that is, if we take the physical - transverse polarization - wave function for one of the
tensor gauge bosons and unphysical - longitudinal polarization - for the second one,
the transition amplitude vanishes.  This Ward identity expresses
the fact that the unphysical - longitudinal polarization states
are not produced in the scattering process. In the fourth and fifth sections the
squared matrix element is calculated together with traces over Dirac and
isotopic matrices for unpolarized particles. In the sixth section we present
the final expression for the cross section (\ref{tensorbosonproductioncrosssection})
and its comparison with the corresponding
cross sections for photons and gluons in QED and QCD.

\section{\it Tensor Gauge Bosons Production Amplitude}

\begin{figure}
\centerline{\hbox{\psfig{figure=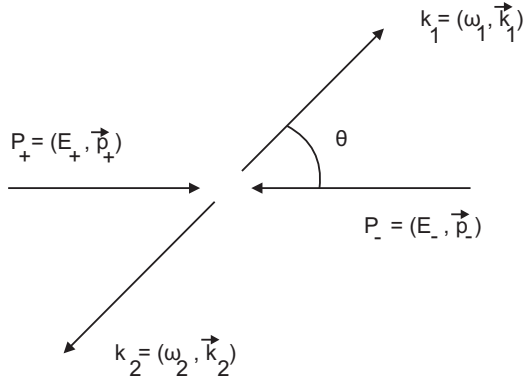,height=5cm,angle=0}}}
\caption[fig1]{The annihilation reaction $f \bar{f}  \rightarrow T T$, shown in the
center-of-mass frame. The $p_{\pm}$ are momenta of the fermions $f \bar{f}$ and $k_{1,2}$
are momenta of the tensor gauge bosons $TT$.}
\label{fig1}
\end{figure}

As we already mentioned in the introduction the process is illustrated in   Fig.\ref{fig1}.
Working in the center-of-mass frame, we make the following assignments:
\be
p_- =(E_-, \vec{p}_-),~~ p_+ =(E_+, \vec{p}_+),~~ k_1 =(\omega_1, \vec{k}_1),~~
k_2 =(\omega_2, \vec{k}_2),
\ee
where $p_{\pm}$ are momenta of the fermions $f^{\pm}$ and $k_{1,2}$
momenta of the tensor gauge
bosons $TT$. All particles are massless $p^{2}_-=p^{2}_+  = k^{2}_1  = k^{2}_2  =0 $.
In the center-of-mass frame
the momenta satisfy the relations $\vec{p}_+  = -\vec{p}_-$, $\vec{k}_2  = -\vec{k}_1$
and $E_- = E_+ = \omega_1 = \omega_2 =E$.
The invariant variables of the process are:
\beqa
s =(p_+ + p_-)^2 = (k_1 + k_2)^2 = 2(p_+ \cdot p_-)  = 2 (k_1 \cdot k_2) \nn\\
t=(p_-  - k_1)^2 = (p_+ - k_2)^2 = -{s\over 2} (1-\cos \theta ) \nn\\
u=(p_-  - k_2)^2 = (p_+ - k_1)^2 = -{s\over 2} (1+\cos \theta ) \nn
\eeqa
where $s= (2E)^2$ and $\theta$ is the scattering angle, so that the
scalar products can be found in the form
\beqa\label{scalarproducts}
 (p_+ \cdot p_-)  =   (k_1 \cdot k_2) = {s\over 2}\nn\\
  (p_-  \cdot  k_1) =  (p_+ \cdot k_2) = {s\over 4} (1-\cos \theta )  \\
 (p_-  \cdot k_2) =  (p_+ \cdot k_1)  = {s\over 4} (1+\cos \theta ). \nn
\eeqa
\begin{figure}
\centerline{\hbox{\psfig{figure=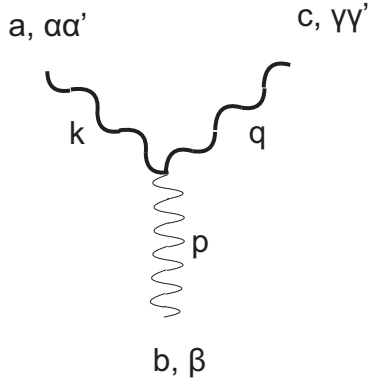,height=5cm,angle=0}}}
\caption[fig1]{The interaction vertex for vector gauge boson V and two
tensor gauge bosons T - the VTT vertex - in non-Abelian tensor gauge
field theory \cite{Savvidy:2005ki}.
Vector gauge bosons are conventionally drawn
as thin wave lines, tensor gauge bosons are thick wave lines.
The Lorentz indices $\alpha\acute{\alpha}$ and momentum $k$ belong to the
first tensor gauge boson, the $\gamma\acute{\gamma}$ and momentum $q$
belong to the second tensor gauge boson, and Lorentz index $\beta$  and
momentum $p$ belong to the vector gauge boson. }
\label{fig2}
\end{figure}

The Feynman rules for this Lagrangian can be derived from the functional
integral over the fermion fields $\psi_{i },~\bar{\psi}_{j },~\psi^{\mu}_{i },~
\bar{\psi}^{\mu}_{j },...$ and over the gauge boson fields
$A^{a}_{\mu},~A^{a}_{\mu\nu},...$. Here Dirac indices are not shown
and $i,j$ and $a$ are indices of the symmetry group G.
In this article we shall ignore subtle aspects
of functional integral quantization procedure simply because we limited
ourselves to calculating only tree diagrams.
Expanding the functional integral in perturbation theory, starting with the free
Lagrangian, at $g=0$, one can see that free theory contains a number of free
fermions of different spins, each of them have equal dimension $d(r)$ of the
representation $r$:~$i,j=1,...,d(r)$ and that
the number of free vector-V and tensor-T gauge bosons
is equal to the number $d(G)$ of generators of the group G:~$a=1,...,d(G)$
\cite{Savvidy:2005fi,Savvidy:2005zm,Savvidy:2005ki}.

In momentum space the interaction vertex of vector gauge boson V with two
tensor gauge bosons T - the VTT vertex - has the form\footnote{See formulas
(62),(65) and (66) in \cite{Savvidy:2005ki} .}
 \cite{Savvidy:2005zm,Savvidy:2005ki}
\be\label{vertexoperator}
V^{abc}_{\alpha\acute{\alpha}\beta\gamma\acute{\gamma}}(k,p,q) =
- g f^{abc} F_{\alpha\acute{\alpha}\beta\gamma\acute{\gamma}} ,
\ee
where
\beqa\label{vertexoperator1}
F_{\alpha\acute{\alpha}\beta\gamma\acute{\gamma}}(k,p,q)
&=& [\eta_{\alpha\beta} (p-k)_{\gamma}+ \eta_{\alpha\gamma} (k-q)_{\beta}
 + \eta_{\beta\gamma} (q-p)_{\alpha}] \eta_{\acute{\alpha}\acute{\gamma}}-\nn\\
-{1\over 2} \{&+&(p-k)_{\gamma}(\eta_{\alpha\acute{\gamma}}
\eta_{\acute{\alpha}\beta}+
\eta_{\alpha\acute{\alpha}} \eta_{\beta\acute{\gamma}})\nn\\
&+& (k-q)_{\beta}(\eta_{\alpha\acute{\gamma}} \eta_{\acute{\alpha}\gamma}+
\eta_{\alpha\acute{\alpha}} \eta_{\gamma\acute{\gamma}})\nn\\
&+& (q-p)_{\alpha} (\eta_{\acute{\alpha}\gamma} \eta_{\beta\acute{\gamma}}+
\eta_{\acute{\alpha}\beta} \eta_{\gamma\acute{\gamma}})\nn\\
&+&(p-k)_{\acute{\alpha}}\eta_{\alpha\beta} \eta_{\gamma\acute{\gamma}}+
(p-k)_{\acute{\gamma}} \eta_{\alpha\beta} \eta_{\acute{\alpha}\gamma}\nn\\
&+&(k-q)_{\acute{\alpha}} \eta_{\alpha\gamma} \eta_{\beta\acute{\gamma}}+
(k-q)_{\acute{\gamma}}\eta_{\alpha\gamma} \eta_{\acute{\alpha}\beta}\nn\\
&+&(q-p)_{\acute{\alpha}} \eta_{\beta\gamma} \eta_{\alpha\acute{\gamma}}+
(q-p)_{\acute{\gamma}}\eta_{\alpha\acute{\alpha}} \eta_{\beta\gamma} \}.
\eeqa
The Lorentz indices $\alpha\acute{\alpha}$ and momentum $k$ belong to the
first tensor gauge boson, the $\gamma\acute{\gamma}$ and momentum $q$
belong to the second tensor gauge boson, and Lorentz index $\beta$  and
momentum $p$ belong to the vector gauge boson. The vertex is shown in
Fig.\ref{fig2}. Vector gauge bosons are conventionally drawn
as thin wave lines, tensor gauge bosons are thick wave lines.

\begin{figure}
\centerline{\hbox{\psfig{figure=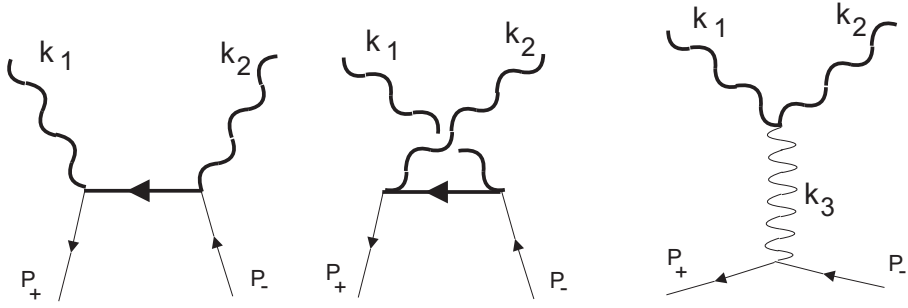,height=4cm,angle=0}}}
\caption[fig1]{Diagrams contributing to fermion-anifermion annihilation to
two tensor gauge bosons. Dirac fermions
are conventionally drawn as thin solid lines, and Rarita-Schwinger spin-vector
fermions by thick solid lines.}
\label{fig3}
\end{figure}

It is convenient to write the differential cross section for our process, with
tensor boson produced into a solid angle $d \Omega$, as
\be
d\sigma = {1 \over 4 (p_+ \cdot p_-)} \vert M \vert^2 d \Phi,
\ee
where the final-state density for two massless tensor gauge bosons is
\beqa
d \Phi = \int {d^3 k_1 \over (2\pi)^3  2 \omega_1}{d^3 k_2 \over (2\pi)^3   2 \omega_2}
(2\pi)^4 \delta(p_+ + p_- - k_1 - k_2)= {1\over 32 \pi^2} d\Omega ,\nn
\eeqa
so that
\be\label{crosssectionformula}
d\sigma = {1 \over 2 s} \vert M \vert^2 {1\over 32 \pi^2} d\Omega.
\ee
We shall calculate the unpolarized cross section for this reaction, to lowest order
in $\alpha = g^2 / 4\pi$. The lowest-order Feynman diagrams contributing to
fermion-antifermion annihilation into a pair of tensor gauge bosons are shown in
Fig.\ref{fig3}. In order $g^2$, there are three diagrams.
Dirac fermions $\psi $
are conventionally drawn as thin solid lines, and Rarita-Schwinger spin-vector
fermions $\psi^{\mu}$ by thick solid lines.
These diagrams are similar
to the QCD diagrams for fermion-antifermion annihilation into a pair of
vector gauge bosons. The difference between these
processes is in the actual expressions for the corresponding interaction vertices.
Thus the probability amplitude of the process can be written in the form
\beqa
i\CM^{\mu\alpha \nu\beta}&=&
(ig)^2 \bar{v}(p_+)\{\gamma^{\mu} t^a {i {1\over 4}g^{\alpha \beta }  \over \not\!p_- -\not\!k_2}t^b \gamma^{\nu}+
\gamma^{\nu}t^b  {i {1\over 4}g^{\beta \alpha }\over \not\!p_-
-\not\!k_1}t^a \gamma^{\mu} - f^{abc} F^{\mu\alpha \rho \nu\beta}
\gamma_{\rho} t^c {1\over k^{2}_{3}}
   \}
u(p_-) \nn
\eeqa
or in equivalent form as
\beqa\label{matrixelement}
 \CM^{\mu\alpha\nu\beta} &=&
(ig)^2 \bar{v}(p_+)\{\gamma^{\mu} t^a {{1\over 4}g^{\alpha \beta } \over \not\!p_-
-\not\!k_2}t^b \gamma^{\nu}+
\gamma^{\nu}t^b  {{1\over 4}g^{\alpha \beta } \over \not\!p_-
-\not\!k_1}t^a \gamma^{\mu} +if^{abc}t^c \gamma_{\rho}{1\over k^{2}_{3}}
F^{\mu\alpha \rho \nu\beta}   \}
u(p_-)\nn \\
\eeqa
where $u(p_-)$ is the wave function of spin $1/2$ fermion and $v(p_+)$ of antifermion.
The Dirac and symmetry group indices are not shown.

\section{\it Gauge Invariance}

Let us check that the amplitude is gauge invariant, that is, if we take
the physical - transverse polarization - wave function $e^T$ for one of the
tensor gauge bosons and longitudinal polarization for the second one $e^L$,
the transition amplitude vanishes $\CM e^T e^L =0$.  This Ward identity expresses
the fact that the unphysical - longitudinal polarization - states
are not produced in the scattering process.

Contracting the matrix element (\ref{matrixelement}) with the  on-shell polarization
tensors of the final tensor gauge bosons $e^{*}_{\mu\alpha}(k_1)$
and $e^{*}_{\nu\beta}(k_2)$ we shall get
\beqa
& \CM^{\mu\alpha\nu\beta}  e^{*}_{\mu\alpha}(k_1)e^{*}_{\nu\beta}(k_2)
= \\
&(ig)^2 \bar{v}(p_+)\{\gamma^{\mu} t^a {{1\over 4}g^{\alpha \beta } \over \not\!p_-
-\not\!k_2}t^b \gamma^{\nu}+
\gamma^{\nu}t^b  {{1\over 4}g^{\alpha \beta } \over \not\!p_-
-\not\!k_1}t^a \gamma^{\mu} +if^{abc}t^c \gamma_{\rho}{1\over k^{2}_{3}}
F^{\mu\alpha \rho \nu\beta}   \}
u(p_-) e^{*}_{\mu\alpha}(k_1)e^{*}_{\nu\beta}(k_2).\nn
\eeqa
Considering the last term
\beqa
if^{abc}t^c ~\bar{v}(p_+)  \gamma_{\rho}u(p_-){1\over k^{2}_{3}}
F^{\mu\alpha \rho \nu\beta}(k_1,k_3,k_2)
e^{*}_{\mu\alpha}(k_1)e^{*}_{\nu\beta}(k_2)  \nn
\eeqa
and taking the polarization tensor  $e^{*}_{\nu\beta}(k_2)$ to be longitudinal
\beqa
e^{*}_{\nu\beta}(k_2)= k_{2\nu} \xi_{\beta} + k_{2\beta} \xi_{\nu} \nn
\eeqa
and the polarization tensor  $e^{*}_{\mu\beta}(k_1)$ to be transversal, we shall get
\beqa
if^{abc}t^c ~\bar{v}(p_+)  \gamma_{\rho}u(p_-){1\over k^{2}_{3}}
F^{\mu\alpha \rho \nu\beta}(k_1,k_3,k_2)
e^{*}_{\mu\alpha}(k_1)(k_{2\nu} \xi_{\beta} + k_{2\beta} \xi_{\nu} ) =\nn\\
=if^{abc}t^c ~\bar{v}(p_+)  \gamma^{\rho}u(p_-){1\over k^{2}_{3}}
[k^{2}_{3} e^{*}_{\rho\alpha}(k_1)
( \xi^{\alpha} -{1\over 2}   \xi^{\alpha} )+
k^{2}_{3} e^{*}_{\alpha\rho}(k_1)
(-{1\over 2} \xi^{\alpha}   + {1\over 4}\xi^{\alpha} )]\nn
\eeqa
and therefore
\beqa\label{lasttermcontribution}
if^{abc}t^c ~\bar{v}(p_+)  \gamma_{\rho}u(p_-){1\over k^{2}_{3}}
F^{\mu\alpha \rho \nu\beta}(k_1,k_3,k_2)
e^{*}_{\mu\alpha}(k_1)(k_{2\nu} \xi_{\beta} + k_{2\beta} \eta_{\nu} ) =\nn\\
=if^{abc}t^c ~\bar{v}(p_+)  \gamma^{\rho}u(p_-)~
{1\over 4}  e^{*}_{\rho\alpha}(k_1)~
 \xi^{\alpha}.
\eeqa
Now let us consider the first two terms
\beqa
  \bar{v}(p_+)\{\gamma^{\mu} t^a {{1\over 4}g^{\alpha \beta } \over \not\!p_-
-\not\!k_2}t^b \gamma^{\nu}+
\gamma^{\nu}t^b  {{1\over 4}g^{\alpha \beta } \over \not\!p_-
-\not\!k_1}t^a \gamma^{\mu}   \} u(p_-)
e^{*}_{\mu\alpha}(k_1)e^{*}_{\nu\beta}(k_2) \nn.
\eeqa
Taking again the polarization tensor to be longitudinal
$
e^{*}_{\nu\beta}(k_2)= k_{2\nu} \xi_{\beta} + k_{2\beta} \xi_{\nu} \nn
$
we shall get
\beqa
  \bar{v}(p_+)\{\gamma^{\mu} t^a {{1\over 4}g^{\alpha \beta } \over \not\!p_-
-\not\!k_2}t^b \gamma^{\nu}+
\gamma^{\nu}t^b  {{1\over 4}g^{\alpha \beta } \over \not\!p_-
-\not\!k_1}t^a \gamma^{\mu}   \} u(p_-)
e^{*}_{\mu\alpha}(k_1)(k_{2\nu} \xi_{\beta} + k_{2\beta} \xi_{\nu}) =\nn \\
= {1\over 4} \bar{v}(p_+)\{- t^a t^b \gamma^{\mu}   +
 t^b t^a  \gamma^{\mu}    \} u(p_-)
e^{*}_{\mu\alpha}(k_1) g^{\alpha \beta } \xi_{\beta} +\nn\\
+ {1\over 4} \bar{v}(p_+)\{\gamma^{\mu} t^a {1 \over \not\!p_-
-\not\!k_2}t^b \gamma^{\nu}+
\gamma^{\nu}t^b  {1 \over \not\!p_-
-\not\!k_1}t^a \gamma^{\mu}   \} u(p_-)
e^{*}_{\mu\alpha}(k_1) g^{\alpha \beta } k_{2\beta} \xi_{\nu}  .\nn
\eeqa
The last term is equal to zero,
$e^{*}_{\mu\alpha}(k_1) g^{\alpha \beta } k_{2\beta}=0$,
see relations (\ref{polarizationidentities}), therefore we have
\beqa
 -{1\over 4}if^{abc} t^c   \bar{v}(p_+) \gamma^{\mu}  u(p_-)
e^{*}_{\mu\alpha}(k_1)   \xi^{\alpha}.
\eeqa
This term precisely cancels the contribution coming from the last term of the
amplitude (\ref{lasttermcontribution}).
Thus the cross term matrix element between transverse and
longitudinal polarizations vanishes $\CM e^T e^L =0$. Our intention now is
to calculate {\it physical matrix element} with two transversal tensor gauge
bosons $\CM e^Te^T$ in the final state.

\section{\it Squared Matrix Element}

The complex conjugate of the scattering amplitude (\ref{matrixelement}) is
\beqa
 \CM^{*\mu\alpha\nu\beta} &=&
(-ig)^2 \bar{u}(p_-)\{\gamma^{\nu} t^b
{{1\over 4}g^{\alpha \beta } \over \not\!p_- -\not\!k_2}t^a \gamma^{\mu}+
\gamma^{\mu}t^a  {{1\over 4}g^{\alpha \beta }
\over \not\!p_- -\not\!k_1}t^b \gamma^{\nu} - if^{abc}t^c \gamma_{\rho}{1\over k^{2}_{3}}
F^{\mu\alpha \rho \nu\beta} \}
v(p_+) \nn
\eeqa
and we can calculate now the squared matrix elements in the form
\beqa
 \CM^{\mu\alpha\nu\beta} \CM^{*\mu^{'}\alpha^{'}\nu^{'}\beta^{'}}&=&
(ig)^2 \bar{v}(p_+)\{\gamma^{\mu} t^a {{1\over 4}g^{\alpha \beta } \over \not\!p_- -\not\!k_2}t^b \gamma^{\nu}+
\gamma^{\nu}t^b  {{1\over 4}g^{\alpha \beta } \over \not\!p_- -\not\!k_1}t^a \gamma^{\mu} +\nn\\
&+&if^{abc}t^c \gamma_{\rho}{1\over k^{2}_{3}}
F^{\mu\alpha \rho \nu\beta} \}
u(p_-)\nn\\
&*&(-ig)^2 \bar{u}(p_-)\{ \gamma^{\nu^{'}} t^{b^{'}} {{1\over 4}g^{\alpha^{'} \beta^{'} } \over \not\!p_- -\not\!k_2}t^{a^{'}} \gamma^{\mu^{'}}+
\gamma^{\mu^{'}}t^{a^{'}}  {{1\over 4}g^{\alpha^{'} \beta^{'} } \over \not\!p_- -\not\!k_1}t^{b^{'}} \gamma^{\nu^{'}} -\nn\\
&-&if^{{a^{'}}{b^{'}}{c^{'}}}t^{c^{'}} \gamma_{\rho^{'}}{1\over k^{2}_{3}}
F^{\mu^{'}\alpha^{'} \rho^{'} \nu^{'}\beta^{'}} \}
v(p_+)\nn
\eeqa
For unpolarized fermions-antifermion scattering the average over the fermion and
antifermion spins is defined as follows:
$$
\vert \CM \vert^2 = {1\over 2}{1\over 2}\sum_{spin ~1/2} \vert M \vert^2,
$$
using completeness relations
$$
\sum_s u^s(p_-)\bar{u}^s(p_-) = \not\!p_- ~, ~~~\sum_s v^s(p_+)\bar{v}^s(p_+) =
\not\!p_+ ~.
$$
Thus averaging over spins of the fermions we shall get
\beqa
 \CM^{\mu\alpha\nu\beta} \CM^{*\mu^{'}\alpha^{'}\nu^{'}\beta^{'}}&=&
{g^4 \over 4} Tr\{ \not\!p_+ [\gamma^{\mu} t^a { {1\over 4}g^{\alpha \beta } \over \not\!p_- -\not\!k_2}t^b \gamma^{\nu}+
\gamma^{\nu}t^b  { {1\over 4} g^{\alpha \beta } \over \not\!p_- -\not\!k_1}t^a \gamma^{\mu} +\nn\\
&+&if^{abc}t^c \gamma_{\rho}{1\over k^{2}_{3}}
F^{\mu\alpha \rho \nu\beta}]
\nn\\
& & \not\!p_-  [ \gamma^{\nu^{'}} t^{b^{'}} {{1\over 4}g^{\alpha^{'} \beta^{'}} \over \not\!p_- -\not\!k_2}t^{a^{'}} \gamma^{\mu^{'}}+
\gamma^{\mu^{'}}t^{a^{'}}  {{1\over 4}g^{\alpha^{'} \beta^{'}} \over \not\!p_- -\not\!k_1}t^{b^{'}} \gamma^{\nu^{'}} -\nn\\
&-&if^{{a^{'}}{b^{'}}{c^{'}}}t^{c^{'}} \gamma_{\rho^{'}}{1\over k^{2}_{3}}
F^{\mu^{'}\alpha^{'} \rho^{'} \nu^{'}\beta^{'}} ] \}
 \nn.
\eeqa
Contracting the last expression with the transversal on-shell  polarization
tensors of the final tensor gauge bosons $e^{*}_{\mu\alpha}(k_1)$
and $e^{*}_{\nu\beta}(k_2)$ we shall get the probability amplitude in the form
\beqa
 &&\CM^{\mu\alpha\nu\beta} \CM^{*\mu^{'}\alpha^{'}\nu^{'}\beta^{'}}
 e^{*}_{\mu\alpha}(k_1)e^{*}_{\nu\beta}(k_2)
e_{\mu^{'}\alpha^{'}}(k_1)e_{\nu^{'}\beta^{'}}(k_2)=\nn\\
&=&{g^4 \over 4} Tr\{ \not\!p_+ [\gamma^{\mu} t^a { {1\over 4}g^{\alpha \beta } \over \not\!p_- -\not\!k_2}t^b \gamma^{\nu}+
\gamma^{\nu}t^b  { {1\over 4} g^{\alpha \beta } \over \not\!p_- -\not\!k_1}t^a \gamma^{\mu} +
if^{abc}t^c \gamma_{\rho}{1\over k^{2}_{3}}
F^{\mu\alpha \rho \nu\beta}]
\nn\\
&& \not\!p_-  [ \gamma^{\nu^{'}} t^{b^{'}} {{1\over 4}g^{\alpha^{'} \beta^{'}} \over \not\!p_- -\not\!k_2}t^{a^{'}} \gamma^{\mu^{'}}+
\gamma^{\mu^{'}}t^{a^{'}}  {{1\over 4}g^{\alpha^{'} \beta^{'}} \over \not\!p_-
-\not\!k_1}t^{b^{'}} \gamma^{\nu^{'}} -if^{{a^{'}}{b^{'}}{c^{'}}}t^{c^{'}} \gamma_{\rho^{'}}{1\over k^{2}_{3}}
F^{\mu^{'}\alpha^{'} \rho^{'} \nu^{'}\beta^{'}} ]  \}\nn\\
&&e^{*}_{\mu\alpha}(k_1)e^{*}_{\nu\beta}(k_2)
e_{\mu^{'}\alpha^{'}}(k_1)e_{\nu^{'}\beta^{'}}(k_2).
 \nn
\eeqa
Using the explicit form of the vertex operator $F^{\mu\alpha \rho \nu\beta}$
(\ref{vertexoperator}),  (\ref{vertexoperator1}) and
the orthogonality properties of the tensor gauge boson wave functions
\beqa\label{polarizationidentities}
k^{\mu}_{1} e_{\mu \alpha}(k_1)=k^{\alpha}_{1} e_{\mu \alpha}(k_1)=
k^{\mu}_{2} e_{\mu \alpha}(k_1)=k^{\alpha}_{2} e_{\mu \alpha}(k_1)=0, \\
k^{\mu}_{2} e_{\mu \alpha}(k_2)=k^{\alpha}_{2} e_{\mu \alpha}(k_2)=
k^{\mu}_{1} e_{\mu \alpha}(k_2)=k^{\alpha}_{1} e_{\mu \alpha}(k_2)=0,\nn
\eeqa
where the last relations follow  from the fact that $\vec{k}_1  \parallel \vec{k}_2$
in the process of Fig.\ref{fig1}, we shall get
\beqa
 &&\CM^{\mu\alpha\nu\beta} \CM^{*\mu^{'}\alpha^{'}\nu^{'}\beta^{'}}
 e^{*}_{\mu\alpha}(k_1)e^{*}_{\nu\beta}(k_2)
e_{\mu^{'}\alpha^{'}}(k_1)e_{\nu^{'}\beta^{'}}(k_2)=\nn\\
&=&{g^4 \over 4} Tr\{ \not\!p_+ [\gamma^{\mu} t^a { {1\over 4}g^{\alpha \beta } \over \not\!p_- -\not\!k_2}t^b \gamma^{\nu}+
\gamma^{\nu}t^b  { {1\over 4} g^{\alpha \beta } \over \not\!p_- -\not\!k_1}t^a \gamma^{\mu} +
if^{abc}t^c \gamma_{\rho}{1\over k^{2}_{3}}
(k_2 -k_1)^{\rho }(g^{\mu \nu} g^{\alpha\beta} -
{1\over 2}g^{\mu\beta } g^{\nu\alpha})]
\nn\\
&& \not\!p_-  [ \gamma^{\nu^{'}} t^{b^{'}} {{1\over 4}g^{\alpha^{'} \beta^{'}} \over \not\!p_- -\not\!k_2}t^{a^{'}} \gamma^{\mu^{'}}+
\gamma^{\mu^{'}}t^{a^{'}}  {{1\over 4}g^{\alpha^{'} \beta^{'}} \over \not\!p_-
-\not\!k_1}t^{b^{'}} \gamma^{\nu^{'}} -if^{{a^{'}}{b^{'}}{c^{'}}}t^{c^{'}} \gamma_{\rho^{'}}{1\over k^{2}_{3}}
(k_2 -k_1)^{\rho{'} }(g^{\mu{'} \nu{'}} g^{\alpha{'}\beta{'}} -
{1\over 2}g^{\mu{'}\beta{'} } g^{\nu{'}\alpha{'}}) ]  \}\nn\\
&&e^{*}_{\mu\alpha}(k_1)e^{*}_{\nu\beta}(k_2)
e_{\mu^{'}\alpha^{'}}(k_1)e_{\nu^{'}\beta^{'}}(k_2).
 \nn
\eeqa
As the next step we shall calculate  the sum over transversal tensor gauge bosons
polarizations.
The sum over transversal polarizations of the helicity-two tensor gauge boson is
\cite{van Dam:1970vg,Savvidy:2005ki}
\beqa
\sum e^{*}_{\mu\alpha}(k_1 )e_{\mu^{'}\alpha^{'}}(k_1) = {1 \over 2}[(-g_{\mu\mu^{'}} +
{ k_{1\mu} \tilde{k}_{1\mu^{'}} + \tilde{k}_{1\mu} k_{1\mu^{'}} \over
k_1 \tilde{k}_1})(-g_{\alpha\alpha^{'}} +
{ k_{1\alpha} \tilde{k}_{1\alpha^{'}} + \tilde{k}_{1\alpha} k_{1\alpha^{'}} \over
k_1 \tilde{k}_1})+\nn\\
+(-g_{\mu\alpha^{'}} +
{ k_{1\mu} \tilde{k}_{1\alpha^{'}} + \tilde{k}_{1\mu} k_{1\alpha^{'}} \over
k_1 \tilde{k}_1})(-g_{\alpha\mu^{'}} +
{ k_{1\alpha} \tilde{k}_{1\mu^{'}} + \tilde{k}_{1\alpha} k_{1\mu^{'}} \over
k_1 \tilde{k}_1})-\nn\\
-(-g_{\mu\alpha} +
{ k_{1\mu} \tilde{k}_{1\alpha} + \tilde{k}_{1\mu} k_{1\alpha} \over
k_1 \tilde{k}_1})(-g_{\mu^{'}\alpha^{'}} +
{ k_{1\mu^{'}} \tilde{k}_{1\alpha^{'}} + \tilde{k}_{1\mu^{'}} k_{1\alpha^{'}} \over
k_1 \tilde{k}_1})]\nn,
\eeqa
where $k_{1\mu} = (\omega_1,\vec{k}_1 ) $ and $\tilde{k}_{1\mu} =
(\omega_1,-\vec{k}_1 ) $. The explicit form of the transversal
polarization tensors, when momentum is aligned along the third axis,
is given by the matrices \cite{van Dam:1970vg,Savvidy:2005ki}
\beqa
e^{1}_{\mu\alpha}={1\over \sqrt{2}}
\left( \begin{array}{llll}
  0,0,~~0,0\\
  0,1,~~0,0\\
  0,0,-1,0\\
  0,0,~~0,0
\end{array} \right), e^{2}_{\mu\alpha}={1\over \sqrt{2}}
\left( \begin{array}{ll}
  0,0,0,0\\
  0,0,1,0\\
  0,1,0,0\\
  0,0,0,0
\end{array} \right).  \nn
\eeqa
From the kinematics of the process in Fig.\ref{fig1}
it follows that $\omega_2 =\omega_1$ and
$\vec{k}_2 = - \vec{k}_1$ therefore
\beqa
\tilde{k}_{1\mu}= k_{2\mu},~~~~ \tilde{k}_{2\mu}= k_{1\mu} \nn
\eeqa
and the average over polarizations can be rewritten as
\beqa\label{sumofpolarization}
\sum e^{*}_{\mu\alpha}(k_1 )e_{\mu^{'}\alpha^{'}}(k_1) =
{1 \over 2}(E_{\mu\mu^{'}} E_{\alpha\alpha^{'}} +
E_{\mu\alpha^{'}}E_{\alpha\mu^{'}} -E_{\mu\alpha}E_{\mu^{'}\alpha^{'}}) ,
\eeqa
where
\beqa
E_{\mu\mu^{'}}= -g_{\mu\mu^{'}} +
{ k_{1\mu} k_{2\mu^{'}} + k_{2\mu} k_{1\mu^{'}} \over
k_1 \cdot k_2}.\nn
\eeqa
Thus the average over tensor gauge boson polarizations gives
\beqa\label{avarageamplitude}
 &&\CM^{\mu\alpha\nu\beta} \CM^{*\mu^{'}\alpha^{'}\nu^{'}\beta^{'}}
 \sum e^{*}_{\mu\alpha}(k_1)e^{*}_{\nu\beta}(k_2)
\sum e_{\mu^{'}\alpha^{'}}(k_1)e_{\nu^{'}\beta^{'}}(k_2)  =\\
&=&{g^4 \over 4} Tr\{ \not\!p_+ [\gamma^{\mu} t^a { {1 \over 4}g^{\alpha \beta } \over \not\!p_- -\not\!k_2}t^b \gamma^{\nu}+
\gamma^{\nu}t^b  {{1 \over 4} g^{\alpha \beta } \over \not\!p_- -\not\!k_1}t^a \gamma^{\mu} +
if^{abc}t^c  {1\over k^{2}_{3}}
(\not\!k_2 -\not\!k_1) (g^{\mu \nu} g^{\alpha\beta} -
{1\over 2}g^{\mu\beta } g^{\nu\alpha})]
\nn\\
&& \not\!p_-  [ \gamma^{\nu^{'}} t^{b^{'}} {{1 \over 4}g^{\alpha^{'} \beta^{'}} \over \not\!p_- -\not\!k_2}t^{a^{'}} \gamma^{\mu^{'}}+
\gamma^{\mu^{'}}t^{a^{'}}  {{1 \over 4}g^{\alpha^{'} \beta^{'}} \over \not\!p_-
-\not\!k_1}t^{b^{'}} \gamma^{\nu^{'}} -if^{{a^{'}}{b^{'}}{c^{'}}}t^{c^{'}}
{1\over k^{2}_{3}}
(\not\!k_2 -\not\!k_1) (g^{\mu{'} \nu{'}} g^{\alpha{'}\beta{'}} -
{1\over 2}g^{\mu{'}\beta{'} } g^{\nu{'}\alpha{'}})  ] \}\nn\\
&&{\delta^{aa^{'}} \over d(r)} {\delta^{bb^{'}} \over d(r)}
{1 \over 2}(E_{\mu\mu^{'}} E_{\alpha\alpha^{'}} +
E_{\mu\alpha^{'}}E_{\alpha\mu^{'}} -E_{\mu\alpha}E_{\mu^{'}\alpha^{'}})
{1 \over 2}(E_{\nu\nu^{'}} E_{\beta\beta^{'}} +
E_{\nu\beta^{'}}E_{\beta\nu^{'}} -E_{\nu\beta}E_{\nu^{'}\beta^{'}}) .\nn
\eeqa
In the next section we shall evaluate these traces and summation over polarizations.

\section{\it  Evaluation of Traces }
In order to evaluate the squared matrix element in the last expression (\ref{avarageamplitude})
we have to calculate traces and then perform summation over polarizations.
We shall use convenient  notations for the different terms in the amplitude.
The whole amplitude will be expressed as a symbolic sum of three terms
$$
\CM = R+L+G,
$$
exactly corresponding to the three Feynman diagrams in Fig.\ref{fig3},
so that the squared amplitude (\ref{avarageamplitude}) shall have nine terms
$$
\CM\CM^* =(R+L+G)(R+L+G)^* .
$$
The first contribution can be evaluated in the following way:
\beqa
(GG^*)^{\mu\alpha\nu\beta~\mu^{'}\alpha^{'}\nu^{'}\beta^{'}} &=&
{g^4 \over 4 d^2(r)} Tr\{ \not\!p_+
if^{abc}t^c  {1\over k^{2}_{3}}
(g^{\mu \nu} g^{\alpha\beta} -
{1\over 2}g^{\mu\beta } g^{\nu\alpha})(\not\!k_2 -\not\!k_1)\nn\\
&& \not\!p_-
(-i)f^{{a^{'}}{b^{'}}{c^{'}}}t^{c^{'}}  {1\over k^{2}_{3}}
(g^{\mu{'} \nu{'}} g^{\alpha{'}\beta{'}} -
{1\over 2}g^{\mu{'}\beta{'} } g^{\nu{'}\alpha{'}}) (\not\!k_2 -\not\!k_1)        \}
 \delta^{aa^{'}}\delta^{bb^{'}}  =
 \nn\\
 &=&   {g^4 \over 4d^2(r)} tr(f^{a  b c }f^{a  b c^{'} } t^{c} t^{c^{'}}   )
{Tr\{ \not\!p_+
(\not\!k_2 -\not\!k_1)
 \not\!p_-
  (\not\!k_2 -\not\!k_1)      \}\over ( 2k_1  k_2)(2k_1 k_2)}\nn\\
  &&
  (g^{\mu \nu} g^{\alpha\beta} -
{1\over 2}g^{\mu\beta } g^{\nu\alpha}) (g^{\mu{'} \nu{'}}g^{\alpha{'}\beta{'}} -
{1\over 2}g^{\mu{'}\beta{'} } g^{\nu{'}\alpha{'}})  \nn.
\eeqa
We can calculate traces over the symmetry group indices using formulas from the Appendix A:
$$
tr(f^{a  b c }f^{a  b c^{'} } t^{c} t^{c^{'}})=d(r) C_{2}(r)C_{2}(G)
= d(G) C(r) C_{2}(G)= {N(N^2 -1) \over 2}
$$
and then the traces of gamma matrices using relation from the Appendix B:
\beqa
 &   {g^4 \over 4d^2(r)} d(r) C_{2}(r) C_{2}(G)
8 {    p_+\cdot (k_2-k_1)~p_-\cdot (k_2-k_1) +   p_+\cdot p_- ~k_1\cdot k_2
\over ( 2k_1  k_2)(2k_1 k_2)}\nn\\
  &
  (g^{\mu \nu} g^{\alpha\beta} -
{1\over 2}g^{\mu\beta } g^{\nu\alpha}) (g^{\mu{'} \nu{'}}g^{\alpha{'}\beta{'}} -
{1\over 2}g^{\mu{'}\beta{'} } g^{\nu{'}\alpha{'}})=\nn\\
 &= {g^4 \over 4d^2(r)}d(r) C_{2}(r) C_{2}(G)
 2  \sin^2\theta
  (g^{\mu \nu} g^{\alpha\beta} -
{1\over 2}g^{\mu\beta } g^{\nu\alpha}) (g^{\mu{'} \nu{'}}g^{\alpha{'}\beta{'}} -
{1\over 2}g^{\mu{'}\beta{'} } g^{\nu{'}\alpha{'}})\nn.
\eeqa
Now it is easy to calculate summation over tensor gauge boson polarizations
using expression (\ref{sumofpolarization}) and the corresponding scalar products
(\ref{scalarproducts})
\beqa\label{1}
GG^* &=&
  {g^4 \over 4d^2(r)}d(r) C_{2}(r) C_{2}(G)~   \sin^2\theta  .
\eeqa
The next contribution in (\ref{avarageamplitude}) can be evaluated as follows:
\beqa
&&(LG^*)^{\mu\alpha\nu\beta~\mu^{'}\alpha^{'}\nu^{'}\beta^{'}} =
{g^4 \over 4d^2(r)} Tr\{ \not\!p_+ [\gamma^{\mu} t^a
{{1 \over 4}g^{\alpha \beta } \over \not\!p_- -\not\!k_2}t^b \gamma^{\nu}  ]
  \nn\\
 &&\not\!p_- [- if^{{a^{'}}{b^{'}}{c^{'}}}t^{c^{'}}  {1\over k^{2}_{3}}
(g^{\mu{'} \nu{'}}g^{\alpha{'}\beta{'}} -
{1\over 2}g^{\mu{'}\beta{'} } g^{\nu{'}\alpha{'}})(\not\!k_2 -\not\!k_1)     ] \}
\delta^{aa^{'}}\delta^{bb^{'}} =\nn\\
&=&-i {g^4 \over 4d^2(r)} tr(f^{a  b c } t^a t^b t^{c} )
{Tr\{ \not\!p_+  \gamma^{\mu}
(\not\!p_- -\not\!k_2)\gamma^{\nu}
 \not\!p_-
  (\not\!k_2 -\not\!k_1)      \}\over (-2p_- k_2)(2k_1 k_2)}
 {1 \over 4}g^{\alpha \beta }(g^{\mu{'} \nu{'}}g^{\alpha{'}\beta{'}} -
{1\over 2}g^{\mu{'}\beta{'} } g^{\nu{'}\alpha{'}})\nn,
\eeqa
and then using traces from the Appendix A and the Appendix B we shall get
\beqa
&{g^4 \over 4d^2(r)}   { d(r)C_{2}(r)C_{2}(G)\over 2}
4\{ g^{\mu\nu } [p_+\cdot p_- ~ k_1 \cdot k_2  + p_+\cdot (k_2 -k_1)~p_-\cdot k_2
+ p_+\cdot k_2 ~ p_-\cdot(k_2 -k_1)]+\nn\\
 &+ k_1\cdot k_2 (p^{\mu}_{+}p^{\nu}_{-} - p^{\mu}_{-}p^{\nu}_{+} )
 + 2p_-\cdot(k_2 -k_1) p^{\mu}_{+}p^{\nu}_{-}
 +2p_+\cdot(k_2 -k_1) p^{\mu}_{-}p^{\nu}_{-}  \}\nn\\
 &{1 \over  (-2p_- k_2)(2k_1 k_2)}
 {1 \over 4} g^{\alpha \beta } (g^{\mu{'} \nu{'}} g^{\alpha{'}\beta{'}} -
{1\over 2} g^{\mu{'}\beta{'} } g^{\nu{'}\alpha{'}} ).\nn
\eeqa
Using again expression (\ref{sumofpolarization}) and scalar products
(\ref{scalarproducts}) we can sum over the polarizations
of tensor gauge bosons:
\beqa\label{2}
LG^* &=&{g^4 \over 4d^2(r)}    d(r)C_{2}(r)C_{2}(G)
(- {1\over 4}  \sin^{2}\theta ).~~~~
\eeqa
The third contribution is
\beqa
&&(RG^*)^{\mu\alpha\nu\beta~\mu^{'}\alpha^{'}\nu^{'}\beta^{'}}=
{g^4 \over 4d^2(r)} Tr\{ \not\!p_+ [\gamma^{\nu}t^b
{{1 \over 4}g^{\alpha \beta } \over \not\!p_- -\not\!k_1}t^a \gamma^{\mu} ] \nn\\
&& \not\!p_-  [ - if^{{a^{'}}{b^{'}}{c^{'}}}t^{c^{'}}
 {1\over k^{2}_{3}}
(g^{\mu{'} \nu{'}}g^{\alpha{'}\beta{'}} -
{1\over 2}g^{\mu{'}\beta{'} } g^{\nu{'}\alpha{'}})( \not\!k_2 - \not\!k_1)      ] \}
\delta^{aa^{'}}\delta^{bb^{'}}  =\nn\\
&=&-i {g^4 \over 4d^2(r)} tr(f^{a  b c } t^b t^a  t^{c} )
{Tr\{ \not\!p_+  \gamma^{\nu}
(\not\!p_- -\not\!k_1)\gamma^{\mu}
 \not\!p_-
  (\not\!k_2 -\not\!k_1)      \}\over (-2p_- k_1)(2k_1 k_2)}
{1 \over 4}g^{\alpha \beta }(g^{\mu{'} \nu{'}}g^{\alpha{'}\beta{'}} -
{1\over 2}g^{\mu{'}\beta{'} } g^{\nu{'}\alpha{'}})\nn
\eeqa
and can be evaluated in the similar way:
\beqa
{g^4 \over 4d^2(r)} (&-&{ d(r)C_{2}(r)C_{2}(G)\over 2})
      4\{ g^{\nu\mu } [-p_+\cdot p_- ~ k_1 \cdot k_2  + p_+\cdot (k_2 -k_1)~p_-\cdot k_1
+ p_+\cdot k_1 ~ p_-\cdot(k_2 -k_1)]+\nn\\
 &+& k_1\cdot k_2 (p^{\nu}_{-} p^{\mu}_{+} - p^{\nu}_{+} p^{\mu}_{-} )
 + 2p_-\cdot(k_2 -k_1) p^{\nu}_{+}p^{\mu}_{-}
 +2p_+\cdot(k_2 -k_1) p^{\nu}_{-} p^{\mu}_{-}  \}
{1 \over  (-2p_- k_1)(2k_1 k_2)} \nn \\
&&{1 \over 4}g^{\alpha \beta }(g^{\mu{'} \nu{'}}g^{\alpha{'}\beta{'}} -
{1\over 2}g^{\mu{'}\beta{'} } g^{\nu{'}\alpha{'}}), \nn
\eeqa
so that after summation over polarization we shall get:
\beqa\label{3}
(RG^*)= {g^4 \over 4d^2(r)}   d(r)C_{2}(r)C_{2}(G) (- {1\over 4} \sin^{2}\theta).
\eeqa
As one can get convinced, the next two terms $GL^*$ and   $GR^*$ give similar
contributions:
\beqa\label{4}
GL^* =  {g^4 \over 4d^2(r)}    d(r)C_{2}(r)C_{2}(G)
(- {1\over 4}  \sin^{2}\theta ),
\eeqa
\beqa\label{5}
 GR^*  =
     {g^4 \over 4d^2(r)}   d(r)C_{2}(r)C_{2}(G) (- {1\over 4} \sin^{2}\theta).
\eeqa
The sixth contribution is
\beqa
&&(LL^*)^{\mu\alpha\nu\beta~\mu^{'}\alpha^{'}\nu^{'}\beta^{'}}={g^4 \over 4d^{2}(r)}
Tr\{ \not\!p_+ [\gamma^{\mu} t^a
{{1 \over 4}g^{\alpha \beta } \over \not\!p_- -\not\!k_2}t^b \gamma^{\nu}]
  \not\!p_-  [\gamma^{\nu^{'}} t^{b^{'}}
{{1 \over 4}g^{\alpha^{'} \beta^{'}} \over \not\!p_- -\not\!k_2}t^{a^{'}} \gamma^{\mu^{'}} ] \}
\delta^{aa^{'}}\delta^{bb^{'}}
=\nn\\
&=&{g^4 \over 4 d^{2}(r)} tr( t^a t^b t^{b }t^{a } )
{Tr \{ \not\!p_+  \gamma^{\mu}
( \not\!p_- -\not\!k_2 ) \gamma^{\nu}
 \not\!p_-   \gamma^{\nu^{'} }
( \not\!p_- -\not\!k_2  )  \gamma^{\mu^{'} }   \}
\over (2p_-k_2) (2p_-k_2)}
{1 \over 4}g^{\alpha \beta } {1 \over 4}g^{\alpha^{'} \beta^{'}}\nn
\eeqa
and involves trace of eight gamma matrices. It can be performed
using expression presented in the Appendix B:
\beqa
&{g^4 \over 4d^{2}(r)} d(r)C_{2}(r)C_{2}(r)
\{16 p^{\nu}_{-} p^{\nu^{'}}_{-} [p^{\mu}_{-}p^{\mu^{'}}_{+}
  + p^{\mu}_{+}  p^{\mu^{'}}_{-}
- p_+\cdot p_- g^{\mu\mu^{'}}] +\nn\\
&+8p^{\nu^{'}}_{-}[(p_+\cdot k_2 ~p^{\mu^{'}}_{-} + p_- \cdot k_2 ~p^{\mu^{'}}_{+} )g^{\mu\nu}-
(p_+ \cdot k_2 ~p^{\mu}_{-} - p_-\cdot k_2~ p^{\mu}_{+} )g^{\mu^{'} \nu}  +
(p_+ \cdot k_2 ~p^{\nu}_{-} - p_- \cdot k_2 ~p^{\nu}_{+} )g^{ \mu  \mu^{'}}  ]\nn\\
&+8p^{\nu}_{-}[(p_+ \cdot k_2 ~p^{\mu}_{-} + p_-\cdot k_2 ~p^{\mu}_{+} )g^{\mu^{'}\nu^{'}}-
(p_+ \cdot k_2 ~p^{\mu^{'}}_{-} - p_- \cdot k_2 ~p^{\mu^{'}}_{+} )g^{\mu \nu^{'}}  +
(p_+ \cdot k_2 ~ p^{\nu^{'}}_{-} - p_-\cdot k_2 ~p^{\nu^{'}}_{+} )g^{\mu \mu^{'} }  ]\nn\\
&+8 p_+\cdot k_2~  p_-\cdot k_2[g^{\mu\nu}g^{\mu^{'}\nu^{'}}-g^{\mu\nu^{'}} g^{\mu^{'}\nu}
+g^{\mu\mu^{'}} g^{\nu\nu^{'}} ]\}
 {1 \over (2p_-k_2) (2p_-k_2)}
 {1 \over 4}g^{\alpha \beta } {1 \over 4}g^{\alpha^{'} \beta^{'}}\nn,
\eeqa
and after summation over polarizations we shall get
\beqa\label{6}
LL^* &=&{g^4 \over 4d^{2}(r)} d(r)C_{2}(r)C_{2}(r)~
{1\over 4} \sin^2\theta .
\eeqa
The seventh contribution is identical with the sixth one and gives
\beqa\label{7}
RR^* &=&{g^4 \over 4d^{2}(r)} d(r)C_{2}(r)C_{2}(r)~
{1\over 4} \sin^2\theta .
\eeqa
The eighth contribution is
\beqa
& (LR^*)^{\mu\alpha\nu\beta~\mu^{'}\alpha^{'}\nu^{'}\beta^{'}}={g^4 \over 4d^{2}(r)} Tr\{ \not\!p_+ [\gamma^{\mu} t^a
{{1 \over 4}g^{\alpha \beta } \over \not\!p_- -\not\!k_2}t^b \gamma^{\nu}]
  \not\!p_-  [\gamma^{\mu^{'}}t^{a^{'}}
  {{1 \over 4}g^{\alpha^{'} \beta^{'}} \over \not\!p_- -\not\!k_1}t^{b^{'}} \gamma^{\nu^{'}} ] \}
\delta^{aa^{'}}\delta^{bb^{'}}
=\nn\\
&={g^4 \over 4 d^{2}(r)} tr( t^a t^b t^{a }t^{b } )
{Tr \{ \not\!p_+  \gamma^{\mu}
( \not\!p_- -\not\!k_2 ) \gamma^{\nu}
 \not\!p_-  \gamma^{\mu^{'} }
( \not\!p_- -\not\!k_1  )   \gamma^{\nu^{'} } \}
\over (2p_-k_2) (2p_-k_1)}{1 \over 4}g^{\alpha \beta }
{1 \over 4}g^{\alpha^{'} \beta^{'}}\nn
\eeqa
and gives
\beqa
&{g^4 \over 4d^{2}(r)} d(r) C_{2}(r)
(C_{2}(r)-{1\over 2}C_2(G))
\{-4 k_1 \cdot k_2~  p_{+} \cdot p_{-} ~  g^{\mu\nu^{'}}   g^{\nu\mu^{'}}
+4 p_{+} \cdot k_{2} ~  p_{-} \cdot k_{1} ~    g^{\mu\nu^{'}} g^{\nu\mu^{'}}+\nn\\
&+ 4 p_{+} \cdot k_{1} ~  p_{-} \cdot k_{2} ~   g^{\mu\nu^{'}} g^{\nu\mu^{'}}-4
 {k_1 \cdot k_2~}   p_{+} \cdot p_{-} ~  g^{\mu\nu}  g^{\mu^{'}\nu^{'}}+  \nn\\
& +  4   p_{+} \cdot k_{2} ~   p_{-} \cdot k_{1} ~  g^{\mu\nu}  g^{\mu^{'}\nu^{'}}+4
 p_{+} \cdot k_{1} ~   p_{-} \cdot k_{2} ~  g^{\mu\nu}  g^{\mu^{'}\nu^{'}}+  \nn\\
&+  4   {k_1 \cdot k_2~}   p_{+} \cdot p_{-} ~  g^{\mu\mu^{'}}  g^{\nu\nu^{'}}-4
 p_{+} \cdot k_{2} ~   p_{-} \cdot k_{1} ~  g^{\mu\mu^{'}}  g^{\nu\nu^{'}}-  \nn\\
& - 4   p_{+} \cdot k_{1} ~   p_{-} \cdot k_{2} ~  g^{\mu\mu^{'}}  g^{\nu\nu^{'}}-4
 {k_1 \cdot k_2~}  g^{\nu\nu^{'}}  p^{\mu^{'}}_{+}  p^{\mu}_{-}+
  4   {k_1 \cdot k_2~}  g^{\mu^{'}\nu^{'}}  p^{\nu}_{+}  p^{\mu}_{-}+4   {k_1 \cdot k_2~}
g^{\nu\mu^{'}}  p^{\nu^{'}}_{+}  p^{\mu}_{-}-  \nn\\
&-  4   {k_1 \cdot k_2~}  g^{\nu\nu^{'}}  p^{\mu}_{+}  p^{\mu^{'}}_{-}+8   p_{-} \cdot k_{2} ~
g^{\nu\nu^{'}}  p^{\mu}_{+}  p^{\mu^{'}}_{-}+
  4   {k_1 \cdot k_2~}  g^{\mu\nu^{'}} p^{\nu}_{+}  p^{\mu^{'}}_{-}-8   p_{-} \cdot k_{2} ~
g^{\mu\nu^{'}} p^{\nu}_{+}  p^{\mu^{'}}_{-}-  \nn\\
&-  4   {k_1 \cdot k_2~}  g^{\mu\nu}  p^{\nu^{'}}_{+}  p^{\mu^{'}}_{-}+8   p_{-} \cdot k_{2} ~
g^{\mu\nu}  p^{\nu^{'}}_{+}  p^{\mu^{'}}_{-}-
  8   p_{+} \cdot k_{2} ~  g^{\nu\nu^{'}}  p^{\mu}_{-}  p^{\mu^{'}}_{-}-4   {k_1 \cdot k_2~}
g^{\mu^{'}\nu^{'}}  p^{\mu}_{+}  p^{\nu}_{-}+  \nn\\
&+  8   p_{-} \cdot k_{1} ~  g^{\mu^{'}\nu^{'}}  p^{\mu}_{+}  p^{\nu}_{-}+4   {k_1 \cdot k_2~}
g^{\mu\nu^{'}} p^{\mu^{'}}_{+}  p^{\nu}_{-}-
  8   p_{-} \cdot k_{1} ~  g^{\mu\nu^{'}} p^{\mu^{'}}_{+}  p^{\nu}_{-}-4   {k_1 \cdot k_2~}
g^{\mu\mu^{'}}  p^{\nu^{'}}_{+}  p^{\nu}_{-}+  \nn\\
&+  8   p_{-} \cdot k_{1} ~  g^{\mu\mu^{'}}  p^{\nu^{'}}_{+}  p^{\nu}_{-}+8   p_{+} \cdot k_{1} ~
g^{\mu^{'}\nu^{'}}  p^{\mu}_{-}  p^{\nu}_{-}+
  8   p_{+} \cdot k_{1} ~  g^{\mu\nu^{'}} p^{\mu^{'}}_{-}  p^{\nu}_{-}+8   p_{+} \cdot k_{2} ~
g^{\mu\nu^{'}} p^{\mu^{'}}_{-}  p^{\nu}_{-}-  \nn\\
&-  16   p_{+} \cdot p_{-} ~  g^{\mu\nu^{'}} p^{\mu^{'}}_{-}  p^{\nu}_{-}+16  p^{\nu^{'}}_{+}
p^{\mu}_{-}  p^{\mu^{'}}_{-}  p^{\nu}_{-}+
  4   {k_1 \cdot k_2~}  g^{\nu\mu^{'}}  p^{\mu}_{+}  p^{\nu^{'}}_{-}+4   {k_1 \cdot k_2~}
g^{\mu\nu}  p^{\mu^{'}}_{+}  p^{\nu^{'}}_{-}-  \nn\\
& - 4   {k_1 \cdot k_2~}  g^{\mu\mu^{'}}  p^{\nu}_{+}  p^{\nu^{'}}_{-}+8   p_{+} \cdot k_{2} ~
g^{\mu\nu}  p^{\mu^{'}}_{-}  p^{\nu^{'}}_{-}-
  8   p_{+} \cdot k_{1} ~  g^{\mu\mu^{'}}  p^{\nu}_{-}  p^{\nu^{'}}_{-}+16  p^{\mu}_{+}
p^{\mu^{'}}_{-}  p^{\nu}_{-}  p^{\nu^{'}}_{-}\}
\nn\\
& {1 \over (2p_-k_2) (2p_-k_1)}
 {1 \over 4}g^{\alpha \beta }
{1 \over 4}g^{\alpha^{'} \beta^{'}}.\nn
\eeqa
After summation over polarizations we shall get
\beqa\label{8}
LR^* &=&{g^4 \over 4d^{2}(r)} d(r) C_{2}(r)
(C_{2}(r)-{1\over 2}C_2(G))~(- {1 \over 4}  \sin^2 \theta)
\eeqa
and for the ninth contribution we shall get identically
\beqa\label{9}
RL^* &=&{g^4 \over 4d^{2}(r)} d(r) C_{2}(r)
(C_{2}(r)-{1\over 2}C_2(G))~(- {1 \over 4}  \sin^2 \theta).
\eeqa

We are now in a position  to calculate the total contribution
to the squared matrix element (\ref{avarageamplitude}). Putting
together all pieces of the squared matrix element
(\ref{1}), (\ref{2}), (\ref{3}), (\ref{4}), (\ref{5}),
(\ref{6}), (\ref{7}), (\ref{8}), (\ref{9}), we finally obtain
\beqa\label{squaredmatrixelenent}
&\CM\CM^* ={g^4 \over 4d^{2}(r)} ~d(r) C_{2}(r)~C_2(G)
 {1\over 4}  ~\sin^{2}\theta.
\eeqa

\section{\it Cross Section}

We can calculate now the leading-order differential cross section for the
tensor gauge bosons production in the annihilation process. This process, as we
discussed in the introduction,
receives contribution from three Feynman diagrams shown in  Fig.\ref{fig3}
and for the unpolarized fermion pairs the squared matrix element was
presented above (\ref{squaredmatrixelenent}).
Plugging everything into our general cross-section formula
(\ref{crosssectionformula}) yields the differential cross section
in the center-of-mass frame:
\beqa
d\sigma &=&{g^4 \over 4d^{2}(r)}~ d(r) C_{2}(r)~C_2(G)
 {1\over 4}     \sin^{2}\theta~
{1\over 2s }~{1\over 32 \pi^2} d\Omega=\nn\\
&=&({g^2 \over  4 \pi  })^2 ~{1\over  s }~{  C_{2}(r) C_2(G)\over  64   d(r) }~
 ~  \sin^{2}\theta~ d\Omega=\nn\\
&=& {\alpha^2 \over  s   }   ~{  C_{2}(r) C_2(G) \over  64   d(r) }~
 ~  \sin^{2}\theta~ d\Omega ,
\eeqa
where
$$
\alpha = {g^2 \over  4 \pi  }.
$$
Thus the unpolarized cross section is
\beqa\label{tensorbosonproductioncrosssection}
 d\sigma   &=& {\alpha^2 \over  s   }   ~
{  C_{2}(r) C_2(G) \over  64   d(r) }
 ~  \sin^{2}\theta~ d\Omega,
\eeqa
where for the $SU(N)$ group we have ${ C_{2}(r) C_2(G) \over  64   d(r) }~
 ~=~{  (N^2 -1) \over  128 N  }$.
This cross section should be compared with the analogous annihilation cross sections
in QED and QCD. Indeed let us compare this result with the electron-positron annihilation
into two photons.
The $e^+ e^- \rightarrow \gamma \gamma$  annihilation cross section \cite{Dirac:1930}
in the high-energy limit is
\beqa\label{gammagamma}
d\sigma_{\gamma\gamma} = {\alpha^2 \over  s}{ 1+\cos^2\theta \over \sin^2\theta} d\Omega
\eeqa
except very small angles of order $m/E$. Angular dependence of cross section
is such that at $\theta=\pi/2$ it has a minimum and then increases for small
angles \cite{Duinker:1981qd,Derrick:1986de}.
The quark pair annihilation cross section into two gluons $q \bar{q} \rightarrow g g$
in the leading order of the strong coupling $\alpha_s$ is
\beqa\label{gluglu}
d\sigma_{gg} = {\alpha^2_s \over  s}{C_2(r)C_2(r) \over   d(r)}[ { 1+\cos^2\theta \over \sin^2\theta }
- {C_{2}(G)  \over 4 C_{2}(r)} (1+    \cos^{2}\theta)   ]d\Omega
\eeqa
and also has minimum at $\theta=\pi/2$ and increases for small
scattering angles  \cite{Abe:1993kb}. The production cross section of tensor
gauge bosons (\ref{tensorbosonproductioncrosssection}) shows dramatically
different behavior with its maximum at
$\theta=\pi/2$ and decrease for small angles.

One of the authors (R.F.) is indebted to the Demokritos National
Research Center for kind hospitality. The work of
(R.F.) was supported by ENRAGE (European Network on Random
Geometry), Marie Curie Research Training Network, contract MRTN-CT-2004-
005616. The work of (G.S.) was partially supported by the EEC Grant
no. MRTN-CT-2004-005616.

\section{\it Appendix A}

The gauge group matrices $t^a$ form a representation  $r$ of the Lie
group G. The matrices $t^a$ obey the algebra $[t^a , t^b ]=i f^{abc} t^c$,
where the structure constants $f^{abc}$ are totally antisymmetric. The
invariant operators $C(r)$ and $C_{2}(r)$ are defined by the equations
$$
 t^a t^b  = C_2(r) 1,~~~~~ tr(t^a t^b) = C(r) \delta^{ab}
$$
and satisfy the relation
$$
C(r) = {d(r)\over d(G)} C_{2}(r),
$$
where $d(r)$ is the dimension of the representation $r$.
By convention the $i$ and $a$ are indices of the symmetry group G.
A number of  fermions $\psi^i$ is equal to the
dimension $d(r)$ of the representation $r$:~$i=1,...,d(r)$.
The number of gauge bosons $A^a$ is equal to the number $d(G)$ of
generators of the group G:~$a=1,...,d(G)$. For the fundamental
$N$ and adjoint $G$ representations of the $SU(N)$ groups  we have
$$
C(N)=1/2,~~~C_{2}(N)={N^2 -1 \over 2N},~~~C(G)=1/2=C_{2}(G)=N.
$$
The traces over symmetry group indices now can be evaluated:
\beqa
-i  tr(f^{a  b c } t^a t^b  t^{c}   )&=&{1\over 2}d(r) C_{2}(r)C_{2}(G)
={1\over 2}d(G) C(r) C_{2}(G)={N(N^2 -1) \over 4},\nn\\
tr(f^{a  b c }f^{a  b c^{'} } t^{c} t^{c^{'}})&=& d(r) C_{2}(r)C_{2}(G)
= d(G) C(r) C_{2}(G)= {N(N^2 -1) \over 2},\nn\\
tr(t^{a}  t^{ b }  t^{b} t^{a} )&=& d(r) C_{2}(r)C_{2}(r)
= d(G) C(r) C_{2}(r)= { (N^2 -1)^2 \over 4N},\nn\\
tr(t^{a}  t^{ b }  t^{a} t^{b} )&=&
d(G) C(r) C_{2}(r)(C_{2}(r)-{1\over 2}C_2(G))=
-{ (N^2 -1) \over 4N}.\nn
\eeqa

\section{\it Appendix B}
In this appendix we shall perform calculation of traces which
appear in the squared matrix element (\ref{avarageamplitude}).
The traces under consideration have terms proportional to the
momentum of the tensor gauge bosons $k^{\mu}_1$ and
$k^{\nu}_2$. These terms can be ignored, because
after contraction with the corresponding transverse wave functions
of the tensor gauge bosons $e_{\mu\alpha}(k_1)$
and $e_{\nu\beta}(k_2)$
they give zero contribution. Therefore we shall calculate the traces
up to the longitudinal terms which are proportional to the
vectors $k^{\mu}_1$ and $k^{\nu}_2$. They are
\beqa
GG \sim Tr\{ \not\!p_+
(\not\!k_2 -\not\!k_1)
 \not\!p_-
  (\not\!k_2 -\not\!k_1)\} =
  8   [ p_+\cdot (k_2-k_1)~p_-\cdot (k_2-k_1) +   p_+\cdot p_- ~k_1\cdot k_2],\nn
\eeqa
\beqa
(LG^*)^{\mu\nu} \sim  &&Tr\{ \not\!p_+  \gamma^{\mu}
(\not\!p_- -\not\!k_2)\gamma^{\nu}
 \not\!p_-
  (\not\!k_2 -\not\!k_1)      \}
 =\nn\\
&=&4\{ g^{\mu\nu } [p_+\cdot p_- ~ k_1 \cdot k_2  + p_+\cdot (k_2 -k_1)~p_-\cdot k_2
+ p_+\cdot k_2 ~ p_-\cdot(k_2 -k_1)]+\nn\\
 &+& k_1\cdot k_2 (p^{\mu}_{+}p^{\nu}_{-} - p^{\mu}_{-}p^{\nu}_{+} )
 + 2p_-\cdot(k_2 -k_1) p^{\mu}_{+}p^{\nu}_{-}
 +2p_+\cdot(k_2 -k_1) p^{\mu}_{-}p^{\nu}_{-}  \},
 \nn
\eeqa
\beqa
(RG^*)^{\mu\nu} \sim&&Tr\{ \not\!p_+  \gamma^{\nu}
(\not\!p_- -\not\!k_1)\gamma^{\mu}
 \not\!p_-
  (\not\!k_2 -\not\!k_1)      \}
 =\nn\\
&=&4\{ g^{\nu\mu } [-p_+\cdot p_- ~ k_1 \cdot k_2  + p_+\cdot (k_2 -k_1)~p_-\cdot k_1
+ p_+\cdot k_1 ~ p_-\cdot(k_2 -k_1)]+\nn\\
 &+& k_1\cdot k_2 (p^{\nu}_{-} p^{\mu}_{+} - p^{\nu}_{+} p^{\mu}_{-} )
 + 2p_-\cdot(k_2 -k_1) p^{\nu}_{+}p^{\mu}_{-}
 +2p_+\cdot(k_2 -k_1) p^{\nu}_{-} p^{\mu}_{-}  \},
 \nn
\eeqa
\beqa
(GL^*)^{\mu\nu} \sim &&Tr\{ \not\!p_+  (\not\!k_2 -\not\!k_1)   \not\!p_-
 \gamma^{\nu}
(\not\!p_- -\not\!k_2)  \gamma^{\mu} \}
 =\nn\\
&=&4\{ g^{\mu\nu } [p_+\cdot p_- ~ k_1 \cdot k_2  + p_+\cdot (k_2 -k_1)~p_-\cdot k_2
+ p_+\cdot k_2 ~ p_-\cdot(k_2 -k_1)]+\nn\\
 &+& k_1\cdot k_2 (p^{\mu}_{+}p^{\nu}_{-} - p^{\mu}_{-}p^{\nu}_{+} )
 + 2p_-\cdot(k_2 -k_1) p^{\mu}_{+}p^{\nu}_{-}    +2p_+\cdot(k_2 -k_1)
 p^{\mu}_{-}p^{\nu}_{-}  \},
 \nn
\eeqa
\beqa
(GR^*)^{\mu\nu} \sim &&Tr\{ \not\!p_+  (\not\!k_2 -\not\!k_1)   \not\!p_-
 \gamma^{\mu}
(\not\!p_- -\not\!k_1)  \gamma^{\nu} \}
 =\nn\\
&=&4\{ g^{\mu\nu } [-p_+\cdot p_- ~ k_1 \cdot k_2  + p_+\cdot (k_2 -k_1)~p_-\cdot k_1
+ p_+\cdot k_1 ~ p_-\cdot(k_2 -k_1)]+\nn\\
 &+& k_1\cdot k_2 (p^{\mu}_{+}p^{\nu}_{-} - p^{\mu}_{-}p^{\nu}_{+} )
 + 2p_-\cdot(k_2 -k_1) p^{\nu}_{+}p^{\mu}_{-}
 +2p_+\cdot(k_2 -k_1) p^{\mu}_{-}p^{\nu}_{-}  \},
 \nn
\eeqa
\beqa
&&(LL^*)^{\mu\nu\mu^{'}\nu^{'}  }  \sim   Tr \{ \not\!p_+  \gamma^{\mu}
( \not\!p_- -\not\!k_2 ) \gamma^{\nu}
 \not\!p_-   \gamma^{\nu^{'} }
( \not\!p_- -\not\!k_2  )  \gamma^{\mu^{'} }   \} =\nn\\
&=&
\{16 p^{\nu}_{-} p^{\nu^{'}}_{-} [p^{\mu}_{-}p^{\mu^{'}}_{+}
  + p^{\mu}_{+}  p^{\mu^{'}}_{-}
- p_+\cdot p_- g^{\mu\mu^{'}}] +\nn\\
&+&8p^{\nu^{'}}_{-}[(p_+\cdot k_2 ~p^{\mu^{'}}_{-} + p_- \cdot k_2 ~p^{\mu^{'}}_{+} )g^{\mu\nu}-
(p_+ \cdot k_2 ~p^{\mu}_{-} - p_-\cdot k_2~ p^{\mu}_{+} )g^{\mu^{'} \nu}  +
(p_+ \cdot k_2 ~p^{\nu}_{-} - p_- \cdot k_2 ~p^{\nu}_{+} )g^{ \mu  \mu^{'}}  ]\nn\\
&+&8p^{\nu}_{-}[(p_+ \cdot k_2 ~p^{\mu}_{-} + p_-\cdot k_2 ~p^{\mu}_{+} )g^{\mu^{'}\nu^{'}}-
(p_+ \cdot k_2 ~p^{\mu^{'}}_{-} - p_- \cdot k_2 ~p^{\mu^{'}}_{+} )g^{\mu \nu^{'}}  +
(p_+ \cdot k_2 ~ p^{\nu^{'}}_{-} - p_-\cdot k_2 ~p^{\nu^{'}}_{+} )g^{\mu \mu^{'} }  ]\nn\\
&+&8 p_+\cdot k_2~  p_-\cdot k_2[g^{\mu\nu}g^{\mu^{'}\nu^{'}}-g^{\mu\nu^{'}} g^{\mu^{'}\nu}
+g^{\mu\mu^{'}} g^{\nu\nu^{'}} ]
                                  \} ,\nn
\eeqa
\beqa
&&(RR^*)^{\mu\nu\mu^{'}\nu^{'}  }  \sim   Tr \{ \not\!p_+  \gamma^{\nu}
( \not\!p_- -\not\!k_1 ) \gamma^{\mu}
 \not\!p_-   \gamma^{\mu^{'} }
( \not\!p_- -\not\!k_1  )  \gamma^{\nu^{'} }   \} =\nn\\
&=&
\{16 p^{\mu}_{-} p^{\mu^{'}}_{-} [p^{\nu}_{+} p^{\nu^{'}}_{-}
  + p^{\nu}_{-}  p^{\nu^{'}}_{+}
- p_+\cdot p_- g^{\nu\nu^{'}}] +\nn\\
&+&8p^{\mu^{'}}_{-}[(p_+\cdot k_1 ~p^{\nu^{'}}_{-} + p_- \cdot k_1 ~p^{\nu^{'}}_{+} )g^{\nu\mu}-
(p_+ \cdot k_1 ~p^{\nu}_{-} - p_-\cdot k_1~ p^{\nu}_{+} )g^{\nu^{'} \mu}  +
(p_+ \cdot k_1 ~p^{\mu}_{-} - p_- \cdot k_1 ~p^{\mu}_{+} )g^{ \nu  \nu^{'}}  ]\nn\\
&+&8p^{\mu}_{-}[(p_+ \cdot k_1 ~p^{\nu}_{-} + p_-\cdot k_1 ~p^{\nu}_{+} )g^{\nu^{'}\mu^{'}}-
(p_+ \cdot k_1 ~p^{\nu^{'}}_{-} - p_- \cdot k_1 ~p^{\nu^{'}}_{+} )g^{\nu \mu^{'}}  +
(p_+ \cdot k_1 ~ p^{\mu^{'}}_{-} - p_-\cdot k_1 ~p^{\mu^{'}}_{+} )g^{\nu \nu^{'} }  ]\nn\\
&+&8 p_+\cdot k_1~  p_-\cdot k_1[g^{\nu\mu}g^{\mu^{'}\nu^{'}}-g^{\nu\mu^{'}} g^{\mu\nu^{'}}
+g^{\nu\nu^{'}} g^{\mu\mu^{'}} ]
                                  \}, \nn
\eeqa
\beqa
 (LR^*)^{\mu\nu\mu^{'}\nu^{'}  }  \sim   Tr \{ \not\!p_+  \gamma^{\mu}
( \not\!p_- -\not\!k_2 ) \gamma^{\nu}
 \not\!p_-   \gamma^{\mu^{'} }
( \not\!p_- -\not\!k_1  )  \gamma^{\nu^{'} }   \} =\nn\\
=-4 k_1 \cdot k_2~  p_{+} \cdot p_{-} ~  g^{\mu\nu^{'}}   g^{\nu\mu^{'}}
+4 p_{+} \cdot k_{2} ~  p_{-} \cdot k_{1} ~    g^{\mu\nu^{'}} g^{\nu\mu^{'}}+\nn\\
 4 p_{+} \cdot k_{1} ~  p_{-} \cdot k_{2} ~   g^{\mu\nu^{'}} g^{\nu\mu^{'}}-4
 {k_1 \cdot k_2~}   p_{+} \cdot p_{-} ~  g^{\mu\nu}  g^{\mu^{'}\nu^{'}}+  \nn\\
   4   p_{+} \cdot k_{2} ~   p_{-} \cdot k_{1} ~  g^{\mu\nu}  g^{\mu^{'}\nu^{'}}+4
 p_{+} \cdot k_{1} ~   p_{-} \cdot k_{2} ~  g^{\mu\nu}  g^{\mu^{'}\nu^{'}}+  \nn\\
  4   {k_1 \cdot k_2~}   p_{+} \cdot p_{-} ~  g^{\mu\mu^{'}}  g^{\nu\nu^{'}}-4
 p_{+} \cdot k_{2} ~   p_{-} \cdot k_{1} ~  g^{\mu\mu^{'}}  g^{\nu\nu^{'}}-  \nn\\
  4   p_{+} \cdot k_{1} ~   p_{-} \cdot k_{2} ~  g^{\mu\mu^{'}}  g^{\nu\nu^{'}}-4
 {k_1 \cdot k_2~}  g^{\nu\nu^{'}}  p^{\mu^{'}}_{+}  p^{\mu}_{-}+  \nn\\
  4   {k_1 \cdot k_2~}  g^{\mu^{'}\nu^{'}}  p^{\nu}_{+}  p^{\mu}_{-}+4   {k_1 \cdot k_2~}
g^{\nu\mu^{'}}  p^{\nu^{'}}_{+}  p^{\mu}_{-}-  \nn\\
  4   {k_1 \cdot k_2~}  g^{\nu\nu^{'}}  p^{\mu}_{+}  p^{\mu^{'}}_{-}+8   p_{-} \cdot k_{2} ~
g^{\nu\nu^{'}}  p^{\mu}_{+}  p^{\mu^{'}}_{-}+  \nn\\
  4   {k_1 \cdot k_2~}  g^{\mu\nu^{'}} p^{\nu}_{+}  p^{\mu^{'}}_{-}-8   p_{-} \cdot k_{2} ~
g^{\mu\nu^{'}} p^{\nu}_{+}  p^{\mu^{'}}_{-}-  \nn\\
  4   {k_1 \cdot k_2~}  g^{\mu\nu}  p^{\nu^{'}}_{+}  p^{\mu^{'}}_{-}+8   p_{-} \cdot k_{2} ~
g^{\mu\nu}  p^{\nu^{'}}_{+}  p^{\mu^{'}}_{-}-  \nn\\
  8   p_{+} \cdot k_{2} ~  g^{\nu\nu^{'}}  p^{\mu}_{-}  p^{\mu^{'}}_{-}-4   {k_1 \cdot k_2~}
g^{\mu^{'}\nu^{'}}  p^{\mu}_{+}  p^{\nu}_{-}+  \nn\\
  8   p_{-} \cdot k_{1} ~  g^{\mu^{'}\nu^{'}}  p^{\mu}_{+}  p^{\nu}_{-}+4   {k_1 \cdot k_2~}
g^{\mu\nu^{'}} p^{\mu^{'}}_{+}  p^{\nu}_{-}-  \nn\\
  8   p_{-} \cdot k_{1} ~  g^{\mu\nu^{'}} p^{\mu^{'}}_{+}  p^{\nu}_{-}-4   {k_1 \cdot k_2~}
g^{\mu\mu^{'}}  p^{\nu^{'}}_{+}  p^{\nu}_{-}+  \nn\\
  8   p_{-} \cdot k_{1} ~  g^{\mu\mu^{'}}  p^{\nu^{'}}_{+}  p^{\nu}_{-}+8   p_{+} \cdot k_{1} ~
g^{\mu^{'}\nu^{'}}  p^{\mu}_{-}  p^{\nu}_{-}+  \nn\\
  8   p_{+} \cdot k_{1} ~  g^{\mu\nu^{'}} p^{\mu^{'}}_{-}  p^{\nu}_{-}+8   p_{+} \cdot k_{2} ~
g^{\mu\nu^{'}} p^{\mu^{'}}_{-}  p^{\nu}_{-}-  \nn\\
  16   p_{+} \cdot p_{-} ~  g^{\mu\nu^{'}} p^{\mu^{'}}_{-}  p^{\nu}_{-}+16  p^{\nu^{'}}_{+}
p^{\mu}_{-}  p^{\mu^{'}}_{-}  p^{\nu}_{-}+  \nn\\
  4   {k_1 \cdot k_2~}  g^{\nu\mu^{'}}  p^{\mu}_{+}  p^{\nu^{'}}_{-}+4   {k_1 \cdot k_2~}
g^{\mu\nu}  p^{\mu^{'}}_{+}  p^{\nu^{'}}_{-}-  \nn\\
  4   {k_1 \cdot k_2~}  g^{\mu\mu^{'}}  p^{\nu}_{+}  p^{\nu^{'}}_{-}+8   p_{+} \cdot k_{2} ~
g^{\mu\nu}  p^{\mu^{'}}_{-}  p^{\nu^{'}}_{-}-  \nn\\
  8   p_{+} \cdot k_{1} ~  g^{\mu\mu^{'}}  p^{\nu}_{-}  p^{\nu^{'}}_{-}+16  p^{\mu}_{+}
p^{\mu^{'}}_{-}  p^{\nu}_{-}  p^{\nu^{'}}_{-},\nn
\eeqa
\beqa
 (RL^*)^{\mu\nu\mu^{'}\nu^{'}  }  \sim   Tr \{ \not\!p_+  \gamma^{\nu}
( \not\!p_- -\not\!k_1 ) \gamma^{\mu}
 \not\!p_-   \gamma^{\nu^{'} }
( \not\!p_- -\not\!k_2  )  \gamma^{\mu^{'} }   \} =\nn\\
=-4   k_{1} \cdot k_{2}~   p_{+} \cdot p_{-}~  g^{\mu \nu^{'} }~  g^{\nu \mu^{'} }~+4   p_{+} \cdot k_{2}~
 p_{-} \cdot k_{1}~  g^{\mu \nu^{'} }~  g^{\nu \mu^{'} }~+  \nn\nn\\
  4   p_{+} \cdot k_{1}~   p_{-} \cdot k_{2}~  g^{\mu \nu^{'} }~  g^{\nu \mu^{'} }~-4
 k_{1} \cdot k_{2}~   p_{+} \cdot p_{-}~  g^{\mu \nu }~  g^{\mu^{'} \nu^{'} }~+  \nn\\
  4   p_{+} \cdot k_{2}~   p_{-} \cdot k_{1}~  g^{\mu \nu }~  g^{\mu^{'} \nu^{'} }~+4
 p_{+} \cdot k_{1}~   p_{-} \cdot k_{2}~  g^{\mu \nu }~  g^{\mu^{'} \nu^{'} }~+  \nn\\
  4   k_{1} \cdot k_{2}~   p_{+} \cdot p_{-}~  g^{\mu \mu^{'} }~  g^{\nu \nu^{'} }~-4
 p_{+} \cdot k_{2}~   p_{-} \cdot k_{1}~  g^{\mu \mu^{'} }~  g^{\nu \nu^{'} }~-  \nn\\
  4   p_{+} \cdot k_{1}~   p_{-} \cdot k_{2}~  g^{\mu \mu^{'} }~  g^{\nu \nu^{'} }~-4
 k_{1} \cdot k_{2}~  g^{\nu \nu^{'} }~  p^{\mu^{'}}_{+}~  p^{\mu}_{-}~+  \nn\\
  8   p_{-} \cdot k_{2}~  g^{\nu \nu^{'} }~  p^{\mu^{'}}_{+}~  p^{\mu}_{-}~-4   k_{1} \cdot k_{2}~
g^{\mu^{'} \nu^{'} }~  p^{\nu}_{+}~  p^{\mu}_{-}~+  \nn\\
  8   p_{-} \cdot k_{2}~  g^{\mu^{'} \nu^{'} }~  p^{\nu}_{+}~  p^{\mu}_{-}~+4   k_{1} \cdot k_{2}~
g^{\nu \mu^{'} }~  p^{\nu^{'}}_{+}~  p^{\mu}_{-}~-  \nn\\
  8   p_{-} \cdot k_{2}~  g^{\nu \mu^{'} }~  p^{\nu^{'}}_{+}~  p^{\mu}_{-}~-4   k_{1} \cdot k_{2}~
g^{\nu \nu^{'} }~  p^{\mu}_{+}~  p^{\mu^{'}}_{-}~+  \nn\\
  4   k_{1} \cdot k_{2}~  g^{\mu \nu^{'} }~  p^{\nu}_{+}~  p^{\mu^{'}}_{-}~+4   k_{1} \cdot k_{2}~
g^{\mu \nu }~  p^{\nu^{'}}_{+}~  p^{\mu^{'}}_{-}~-  \nn\\
  8   p_{+} \cdot k_{2}~  g^{\nu \nu^{'} }~  p^{\mu}_{-}~  p^{\mu^{'}}_{-}~+4   k_{1} \cdot k_{2}~
g^{\mu^{'} \nu^{'} }~  p^{\mu}_{+}~  p^{\nu}_{-}~~+  \nn\\
  4   k_{1} \cdot k_{2}~  g^{\mu \nu^{'} }~  p^{\mu^{'}}_{+}~  p^{\nu}_{-}~~-4   k_{1} \cdot k_{2}~
g^{\mu \mu^{'} }~  p^{\nu^{'}}_{+}~  p^{\nu}_{-}~~+  \nn\\
  8   p_{+} \cdot k_{2}~  g^{\mu^{'} \nu^{'} }~  p^{\mu}_{-}~  p^{\nu}_{-}~~+4   k_{1} \cdot k_{2}~
g^{\nu \mu^{'} }~  p^{\mu}_{+}~  p^{\nu^{'}}_{-}~-  \nn\\
  8   p_{-} \cdot k_{1}~  g^{\nu \mu^{'} }~  p^{\mu}_{+}~  p^{\nu^{'}}_{-}~-4   k_{1} \cdot k_{2}~
g^{\mu \nu }~  p^{\mu^{'}}_{+}~  p^{\nu^{'}}_{-}~+  \nn\\
  8   p_{-} \cdot k_{1}~  g^{\mu \nu }~  p^{\mu^{'}}_{+}~  p^{\nu^{'}}_{-}~-4   k_{1} \cdot k_{2}~
g^{\mu \mu^{'} }~  p^{\nu}_{+}~  p^{\nu^{'}}_{-}~+  \nn\\
  8   p_{-} \cdot k_{1}~  g^{\mu \mu^{'} }~  p^{\nu}_{+}~  p^{\nu^{'}}_{-}~+8   p_{+} \cdot k_{1}~
g^{\nu \mu^{'} }~  p^{\mu}_{-}~  p^{\nu^{'}}_{-}~+  \nn\\
  8   p_{+} \cdot k_{2}~  g^{\nu \mu^{'} }~  p^{\mu}_{-}~  p^{\nu^{'}}_{-}~-16   p_{+} \cdot p_{-}~
g^{\nu \mu^{'} }~  p^{\mu}_{-}~  p^{\nu^{'}}_{-}~+  \nn\\
  8   p_{+} \cdot k_{1}~  g^{\mu \nu }~  p^{\mu^{'}}_{-}~  p^{\nu^{'}}_{-}~+16  p^{\nu}_{+}~
p^{\mu}_{-}~  p^{\mu^{'}}_{-}~  p^{\nu^{'}}_{-}~-  \nn\\
  8   p_{+} \cdot k_{1}~  g^{\mu \mu^{'} }~  p^{\nu}_{-}~~  p^{\nu^{'}}_{-}~+16  p^{\mu^{'}}_{+}~
p^{\mu}_{-}~  p^{\nu}_{-}~~  p^{\nu^{'}}_{-}~.\nn\\
\eeqa
All these traces have been calculated with the use of the Mathematica program \cite{Gran}.

\vfill
\end{document}